\documentclass[12pt]{article}
\usepackage{amsmath,color}
\usepackage[dvips]{graphicx}
\input amssym
\input amssym.def

\def\red#1{{\color{red} #1}}
\begin{document}

\def\prg#1{\medskip\noindent{\bf #1}}  \def\ra{\rightarrow}
\def\lra{\leftrightarrow}              \def\Ra{\Rightarrow}
\def\nin{\noindent}                    \def\pd{\partial}
\def\dis{\displaystyle}                \def\inn{\hook}
\def\grl{{GR$_\Lambda$}}               \def\Lra{{\Leftrightarrow}}
\def\cs{{\scriptstyle\rm CS}}          \def\ads3{{\rm AdS$_3$}}
\def\Leff{\hbox{$\mit\L_{\hspace{.6pt}\rm eff}\,$}}
\def\bull{\raise.25ex\hbox{\vrule height.8ex width.8ex}}
\def\ric{{Ric}}                      \def\tric{{(\widetilde{Ric})}}
\def\tmgl{\hbox{TMG$_\Lambda$}}
\def\Lie{{\cal L}\hspace{-.7em}\raise.25ex\hbox{--}\hspace{.2em}}
\def\sS{\hspace{2pt}S\hspace{-0.83em}\diagup}   \def\hd{{^\star}}
\def\dis{\displaystyle}                 \def\ul#1{\underline{#1}}
\def\mb#1{\hbox{{\boldmath $#1$}}}     \def\tgr{{GR$_\parallel$}}
\def\irr#1{^{(#1)}\hspace{-2pt}}

\def\hook{\hbox{\vrule height0pt width4pt depth0.3pt
\vrule height7pt width0.3pt depth0.3pt
\vrule height0pt width2pt depth0pt}\hspace{0.8pt}}
\def\semidirect{\;{\rlap{$\supset$}\times}\;}
\def\first{\rm (1ST)}       \def\second{\hspace{-1cm}\rm (2ND)}
\def\bm#1{\hbox{{\boldmath $#1$}}}
\def\nb#1{\marginpar{{\large\bf #1}}}
\def\ir#1{\,{}^{(#1)}\hspace{-2pt}}   \def\orth{{\perp}}

\def\G{\Gamma}        \def\S{\Sigma}        \def\L{{\mit\Lambda}}
\def\D{\Delta}        \def\Th{\Theta}
\def\a{\alpha}        \def\b{\beta}         \def\g{\gamma}
\def\d{\delta}        \def\m{\mu}           \def\n{\nu}
\def\th{\theta}       \def\k{\kappa}        \def\l{\lambda}
\def\vphi{\varphi}    \def\ve{\varepsilon}  \def\p{\pi}
\def\r{\rho}          \def\Om{\Omega}       \def\om{\omega}
\def\s{\sigma}        \def\t{\tau}          \def\eps{\epsilon}
\def\nab{\nabla}      \def\btz{{\rm BTZ}}   \def\heps{\hat\eps}
\def\bu{{\bar u}}     \def\bv{{\bar v}}     \def\bs{{\bar s}}
\def\bx{{\bar x}}     \def\by{{\bar y}}     \def\bom{{\bar\om}}
\def\tphi{{\tilde\vphi}}  \def\tt{{\tilde t}} \def\bR{{\bar R}}
\def\tG{{\tilde G}}   \def\cF{{\cal F}}      \def\bH{{\bar H}}
\def\cL{{\cal L}}     \def\cM{{\cal M }}     \def\cE{{\cal E}}
\def\cH{{\cal H}}     \def\hcH{\hat{\cH}}    \def\cP{{\cal P}}
\def\cK{{\cal K}}     \def\hcK{\hat{\cK}}    \def\cT{{\cal T}}
\def\cO{{\cal O}}     \def\hcO{\hat{\cal O}} \def\cV{{\cal V}}
\def\tom{{\tilde\omega}}                     \def\cE{{\cal E}}
\def\cR{{\cal R}}    \def\hcT{{\hat\cT}{}}   \def\hcR{{\hat\cR}{}}
\def\tb{{\tilde b}}  \def\tI{{\tilde I}{}}     \def\tv{{\tilde v}}
\def\tT{{\tilde T}}  \def\tR{{\tilde R}}     \def\tcL{{\tilde\cL}}
\def\bA{{\bar A}}     \def\bB{{\bar B}}      \def\bC{{\bar C}}
\def\bG{{\bar G}}     \def\bD{{\bar D}}      \def\bH{{\bar H}}
\def\bK{{\bar K}}     \def\bL{{\bf L}}       \def\cB{{\cal B}}
\def\bj{{\bar j}}     \def\bk{{\bar k}}      \def\bb{{\bar b}}
\def\bm{{\bar m}}     \def\bcE=\mb{\cE}      \def\vth{{\vartheta}}

\def\rdc#1{\hfill\hbox{{\small\texttt{reduce: #1}}}}
\def\chm{\checkmark}  \def\chmr{\red{\chm}}
\def\E{\Bbb{E}}      \def\B{\Bbb{B}}  \def\rn{RN{}}

\def\nn{\nonumber}                    \def\vsm{\vspace{-9pt}}
\def\be{\begin{equation}}             \def\ee{\end{equation}}
\def\ba#1{\begin{array}{#1}}          \def\ea{\end{array}}
\def\bea{\begin{eqnarray} }           \def\eea{\end{eqnarray} }
\def\beann{\begin{eqnarray*} }        \def\eeann{\end{eqnarray*} }
\def\beal{\begin{eqalign}}            \def\eeal{\end{eqalign}}
\def\lab#1{\label{eq:#1}}             \def\eq#1{(\ref{eq:#1})}
\def\bsubeq{\begin{subequations}}     \def\esubeq{\end{subequations}}
\def\bitem{\begin{itemize}}           \def\eitem{\end{itemize}}
\renewcommand{\theequation}{\thesection.\arabic{equation}}
\def\aff#1{{\normalsize #1}}
\title{Entropy of Reissner-Nordstr\"om-like black holes}

\author{M. Blagojevi\'c and B. Cvetkovi\'c\footnote{
        Email addresses: \texttt{mb@ipb.ac.rs, cbranislav@ipb.ac.rs}} \\
\aff{Institute of Physics, University of Belgrade,
                      Pregrevica 118, 11080 Belgrade, Serbia} }
\date{\today}
\maketitle

\begin{abstract}
In Poincar\'e gauge theory, black hole entropy is defined canonically by the variation of a boundary term $\G_H$, located at horizon. For a class of static and spherically symmetric black holes in vacuum, the explicit formula reads $\d\G_H=T\d S$, where $T$ is black hole temperature and $S$ entropy. Here, we analyze a new member of the same class, the Reissner-Nordstr\"om-like black hole with torsion \cite{x1}, where the electric charge of matter is replaced by a gravitational parameter, induced by the existence of torsion. This parameter affects $\d\G_H$ in a way that ensures the validity of the first law.

\end{abstract}

\section{Introduction}
\setcounter{equation}{0}

The discovery of the thermodynamic behavior of black holes in general relativity (GR) has deeply affected our understanding of both classical and quantum aspects of the gravitational dynamics. At the heart of that behavior lies the  concept of entropy, which is classically represented by a boundary term $\G_H$, interpreted as the Noether charge on black hole horizon \cite{x2}. For a class of spherically symmetric black holes in GR without external fields, $\d\G_H$ has the standard form $T\d S$, where $T$ is the black hole temperature and $S$ entropy, equal to one-fourth of the horizon area. This result remains essentially valid (up to a multiplicative constant) for diffeomorphism invariant Lagrangian theories in Riemannian theories of gravity \cite{x2,x3}.
In order to have a deeper understanding of entropy, one is naturally led to investigate to what extent it might depend on the structure of black holes and the type of the gravitational dynamics \cite{x4}.

Entropy and the asymptotic charges (such as energy or angular momentum) are strongly interrelated through the first law of black hole thermodynamics. Recently, a  Hamiltonian approach to entropy was proposed in Ref. \cite{x5}, in which the asymptotic charges and entropy are described as the canonical charges at infinity and horizon, respectively. This approach was primarily intended to describe entropy in the framework of Poincar\'e gauge theory (PG) \cite{x6,x7,x8}, where both the curvature and the torsion are essential ingredients of the new gravitational dynamics. However, for a number of black holes in PG without matter, including Kerr-AdS black holes \emph{with or without torsion} \cite{x9}, it was found, somewhat unexpectedly, that entropy retains its area form (up to a multiplicative factor), and the first law remains valid.

Are there black holes for which entropy and/or the first law change their standard forms? There are two 3-dimensional PG models which are interesting in this respect. First, entropy of the Banados-Teitelboim-Zanelli black hole with torsion is determined by a parameter that measures the strength of torsion \cite{x10}, and second, entropy of the Oliva-Tempo-Troncoso black hole depends on a ``hair" parameter appearing in the metric \cite{x11}. In both cases, the first law remains unchanged. These examples motivate us to explore thermodynamic aspects of a new black hole with torsion, found in the four-dimensional PG by Cembranos and Valcarcel \cite{x1}, see also \cite{x12}. It represents a \emph{gravitational analogue} of the standard Reissner-Nordstr\"om (\rn) black hole, in which the ``electric charge" is produced not by an external source, as in Einstein-Maxwell theory \cite{x13}, but by the \emph{gravitational field in vacuum.}

The paper is organized as follows. In section \ref{sec2}, we present basic elements of the PG dynamics and the Hamiltonian approach to entropy. In section \ref{sec3}, we analyze geometric aspects of the \rn-like black hole in the absence of external fields. The main results of the paper, the derivation of the expression for entropy and the form of the first law, are presented in section \ref{sec4}, and section \ref{sec5} is devoted to concluding remarks.

Our conventions are the same as in Refs. \cite{x9}. The Latin indices $(i,j,\dots)$ are the local Lorentz indices, the Greek indices $(\m,\n,\dots)$ are the coordinate indices, and both run over $0, 1, 2, 3$. The orthonormal coframe (tetrad) 1-form is $\vth^i$, $h_i$ is the dual basis (frame), the interior product of $\vth^k$ with $h_i$ is defined by $h_i\inn \th^k=\d_i^k$, and $\om^{ij}$ is a metric compatible (Lorentz) connection 1-form. The metric components  $\eta_{ij}= (1,-1,-1,-1)$ refer to the local Lorentz basis, the Hodge dual of a form $\a$ is denoted by $\hd\a$, and the wedge product of forms is implicitly understood.

\section{Entropy as the canonical charge}\label{sec2}
\setcounter{equation}{0}

In this chapter, we present a short account of PG and the Hamiltonian approach to entropy.

Poincar\'e gauge theory is a gauge theory of gravity, based on the localization of the Poincar\'e group of spacetime symmetries \cite{x6,x7}. In this process, the tetrad field $\vth^i$ and the Lorentz connection $\om^{ij}$ are found to be the gauge potentials, the related field strengths are identified with the torsion $T^i=d\vth^i+\om^i{}_k\vth^k$ and the curvature $R^{ij}=d\om^{ij}+\om^i{}_k\om^{kj}$, and the spacetime is characterized by a Riemann-Cartan geometry.

In the absence of matter fields, the PG dynamics is completely determined by the gra\-vi\-tational Lagrangian $L_G=L_G(\vth^i,T^i,R^{ij})$ (4-form), which is assumed to be parity invariant and at most quadratic in the field strengths,
\be
L_G=-\hd(a_0R+2\L)+T^i\sum_{n=1}^3\hd(a_n\ir{n}T_i)
             +\frac{1}{2}R^{ij}\sum_{n=1}^6\hd(b_n\ir{n}R_{ij})\,.   \lab{2.1}
\ee
Here,  $\ir{n}T^i$ and $\ir{n}R^{ij}$ are irreducible parts of the field strengths, for details see \cite{x5}, and  $(\L,a_0,a_n,b_n)$ are the gravitational coupling constants. The gravitational field equations, obtained by varying $L_G$ with respect to $\vth^i$ and $\om^{ij}$, can be written in a compact form as
\bsubeq\lab{2.2}
\bea
\d\vth^i:&&\quad \nab H_i+E_i=0\, ,                                  \lab{2.2a}\\
\d\om^{ij}:&&\quad \nab H_{ij}+E_{ij}=0\,,                           \lab{2.2b}
\eea
\esubeq
where $H_i:=\pd L_G/\pd T^i$ and $H_{ij}:=\pd L_G/\pd R^{ij}$ are the covariant momenta (2-forms), and $E_i:=\pd L_g/\pd\vth^i$ and $E_{ij}:=\pd L_G/\pd\om^{ij}$ are the energy-momentum and spin currents (3-forms), respectively.
As we shall see, the covariant momenta play a crucial role in the Hamiltonian approach to black hole entropy.

In the early 1970s, Regge and Teitelboim \cite{x14} demonstrated that the canonical analysis of GR requires the standard ADM Hamiltonian to be modified by adding suitable \emph{boundary terms} at infinity. The role of these terms is twofold: first, they are necessary to ensure the consistency of the variational formalism (the existence of the field equations), and second, their values are associated to the asymptotic conserved charges (energy-momentum or angular momentum). For an early extension of these ideas to PG, see Refs. \cite{x15}, whereas Ref. \cite{x16} offers a comprehensive presentation of the covariant Hamiltonian formalism.

Further generalization of these ideas, based on the general Hamiltonian structure of PG, was focussed on understanding entropy as an additional conserved charge at horizon \cite{x5}. Let $\S$ be a three-dimensional subspace of a stationary black hole spacetime, such that its boundary has two components, one at infinity, $S_\infty$, and the other at horizon, $S_H$.\footnote{The same boundary structure was used by Wald \cite{x2} to study gravitational theories in asymptotically flat Riemannian spacetimes, based on diffeomorphism invariant Lagrangians and the Noether charge technique.} Then, for any Killing vector $\xi$, the associated boundary terms $\G_\infty$ and $\G_H$ are determined by the following variational equations:
\bsubeq\lab{2.3}
\bea
&&\d\G_\infty=\oint_{S_\infty}\d B(\xi)\,,\qquad
       \d\G_H=\oint_{S_H} \d B(\xi)\,,                                  \\
&&\d B(\xi):=(\xi\inn\vth^{i})\d H_i+\d\vth^i(\xi\inn H_i)
   +\frac{1}{2}(\xi\inn\om^{ij})\d H_{ij}
   +\frac{1}{2}\d\om^{ij}(\xi\inn\d H_{ij})\, .
\eea
\esubeq
Here, $\G_\infty$ is a direct extension of the Regge-Teitelboim construction to PG, whereas $\G_H$ defines an additional contribution associated to entropy. The variation $\d$ is required to satisfy the following rules:
\bitem
\item[(r1)] On $S_\infty$, the variation $\d$ acts on the parameters of a black hole solution, but not on the para\-me\-ters of the background configuration.\vsm
\item[(r2)] On $S_H$, the variation $\d$ must keep surface gravity constant.
\eitem
When the asymptotic conditions allow the solutions for $\G_\infty$ and $\G_H$ to exist and be finite,\footnote{The existence of the solutions $\G_\infty$ and $\G_H$ of the variational equations \eq{2.3} will be referred to as $\d$-integrability.} they define the asymptotic charges and black hole entropy, respectively.

Technically, Eqs. \eq{2.3} are derived by requiring that the canonical gauge generator $G$ is a regular (differentiable) functional on the phase space with given asymptotic conditions. The regularity of $G$ is ensured by the relation
\be
\d\G\equiv\d\G_\infty-\d\G_H=0\,,                                    \lab{2.4}
\ee
which is interpreted as the first law of black hole thermodynamics.

\section{\rn-like black holes}\label{sec3}
\setcounter{equation}{0}

Classical solutions of PG help us to better understand the influence of torsion on the gravitational dynamics \cite{x6,x7}. In this section, we focus our attention on the static and spherically symmetric solution found by Cembranos and Valcarcel \cite{x1}, in which the electric charge of the standard \rn\ metric is ``imitated" by a torsion-induced parameter.

\subsection{Geometry}

The metric of asymptotically flat \rn-like black holes has the form
\be
ds^2=N^2 dt^2-\frac{dr^2}{N^2}-r^2(d\th^2+\sin^2\th d\vphi^2)\,,\qquad
N^2=1-\frac{2m}{r}+\frac{q}{r^2}\,,                                  \lab{3.1}
\ee
where $m$ is the usual mass parameter of the Schwarzschild metric, but $q$ differs from the standard electric charge in the \rn\ metric. As we shall see in the next subsection, $q$ will be determined on shell by a particular torsion parameter.

Many geometric aspects of the solution follow directly from the metric. Thus, the outer horizon is located at the larger root of $N^2=0$,
\be
r_+^2-2mr_++q=0\,,\qquad r_{\pm}=m\pm\sqrt{m^2-q}\,.                 \lab{3.2}
\ee
The standard \rn\ black holes are characterized by $q>0$, with two subcases:
$q\le m^2$, where both roots are nonnegative, and $q>m^2$, where the roots $r_\pm$ are complex and there are no horizons.\footnote{In that case, the quadratic field strength singularities found in \eq{3.9} are naked.}. The option $q<0$ is possible only for \rn-like black holes.\footnote{The special limit $q=0$ will be discussed in subsection 3.3.} Further analysis of these puzzling aspects of the horizon structure will be given in the next subsection.

The surface gravity and black hole temperature are given by
\be
\k:=\frac{1}{2}\pd_r N^2\Big|_{r_+}
   =\frac{mr_+-q}{r_+^3}=\frac{r_+^2-q}{2r_+^3}\,,\qquad T:=\frac{\k}{2\pi}\,.
                                                                     \lab{3.3}
\ee
The orthonormal tetrad associated to the metric \eq{3.1} is chosen in the form
\be
\vth^0=N dt\, ,\qquad \vth^1=\frac{dr}{N}\, ,\qquad \vth^2=rd\th\, ,
            \qquad\vth^3=r\sin\th d\vphi\,.                          \lab{3.4}
\ee
Then, a simple calculation of the horizon area yields
\be
A_H=\int_H\vth^2\vth^3=4\pi r_+^2\,.
\ee

To complete the gravitational description of \rn-like black holes in PG, one also needs to specify the Lorentz connection $\om^{ij}$. We start the construction of $\om^{ij}$ by introducing the static and rotationally invariant ansatz for torsion \cite{x1},
\bsubeq
\bea
&&T^0=T^1=A\vth^0\vth^1\, ,                                          \nn\\
&&T^2=-G\vth^-\vth^2- H\vth^-\vth^3\, ,                              \nn\\
&&T^3=-G\vth^-\vth^3+ H\vth^-\vth^2\, ,
\eea
where $\vth^-:=\vth^0-\vth^1$. The torsion functions $A,G$ and $H$ have the forms
\be
A:=-\pd_r N\,,\qquad G=\frac{N}{2r}\, , \qquad H:=\frac{p}{rN}\,,
\ee
\esubeq
where $p$ is a new parameter, the torsion counterpart of the metric parameter $q$. All irreducible parts of $T^i$ are nonvanishing. From the adopted ansatz for torsion, one can calculate the connection $\om^{ij}$ from the implicit formula $d\vth^i+\om^i{}_m\vth^m=T^i$,
\bea
&&\om^{01}=-(\pd_r N)\vth^1\,,                                       \nn\\
&&\om^{0c}=-\om^{1c}=-\frac{N}{2r}\vth^c\,,\qquad c=(2,3)\,,         \nn\\
&&\om^{23}=\frac{\cos\th}{r\sin\th}\vth^3+\frac{p}{Nr}\vth^-\,.      \lab{3.7}
\eea
Then, the nonvanishing components of the RC curvature $R^{ij}=d\om^{ij}+\om^i{}_m\om^{mj}$ are
\bea
&&R_{A2}=-\frac{p}{2r^3}\vth^-\vth^3\,,               \nn\\
&&R_{A3}=\frac{p}{2r^3}\vth^-\vth^2\,,                \nn\\
&&R_{23}=\frac{1}{r^2}\big(p\vth^0\vth^1-\vth^2\vth^3\big)\,,
\eea
where $A=(0,1)$. All irreducible parts of the curvature are nonvanishing.

The above expressions for the pair ($\vth^i,\om^{ij}$), characterized by the  parameters $q$ and $p$, represent a complete ansatz for \rn-like black holes as vacuum solutions of PG.

The PG field strengths, the torsion and the curvature, define three topological invariants (Euler, Pontryagin and Nieh-Yan):
\bea
&&I_E=\ve_{ijmn}R^{mn}R^{ij}=0\, ,\qquad
  I_P=R^{ij}R_{ij}=-\frac{4p}{r^4}\,\heps\, ,                        \nn\\
&&I_{NY}=T^iT_i-R_{ij}\vth^i\vth^j=\frac{2p}{r^2}\,\heps\,,          \lab{3.9}
\eea
where $\heps:=\vth^0\vth^1\vth^2\vth^2$ is the volume 4-form. Thus, \rn-like black holes are singular at $r=0$, and moreover, we have here an interesting topological interpretation of the parameter $p$.

\subsection{Solving the field equations}

The adopted expressions for the pair $(\vth^i,\om^{ij})$ satisfy the PG field equations \eq{2.2} provided the following restrictions on the Lagrangian parameters in \eq{2.1} are imposed:
\bea
&&\L=0\,,\qquad\nn\\
&& a_1-a_0=0\,,\quad a_2+2a_0=0\,,\quad a_3+a_0/2=0\,,               \nn\\
&&b_1=b_4=b_6=0\,,\qquad b_3=-b_2\, ,\qquad b_5=-b_2/3\,.            \lab{3.10}
\eea
In addition to that, the black hole parameters $q$ and $p$ become dynamically interrelated,
\be
q=k p^2\, ,\qquad k:=-\frac{b_2}{3a_0}\, .                           \lab{3.11}
\ee
As a consequence of \eq{3.10},  the general Lagrangian is simplified,
\bea\lab{3.12}
L_G&=&-\hd(a_0R)
    +a_0T^i\hd\Big(\ir{1}T_1-2\ir{2}T_i-\frac{1}{2}\ir{3}T_i\Big)    \nn\\
  &&+\frac{1}{2}b_2R^{ij}\hd\Big(\ir{2}R_{ij}-\ir{3}R_{ij}
                                   -\frac{1}{3}\ir{5}R_{ij}\Big)\,,
\eea
where $a_0$ is taken to be positive, to ensure a smooth transition to GR.

Formula \eq{3.11} clarifies the influence of torsion on the structure of horizons. Note that, according to the Hamiltonian analysis of the particle spectrum of the general PG \cite{y1}, a necessary condition for the gravitational Hamiltonian of the spin 1 sector to be positive is given by  $(b_2+b_5>0$, $b_4+b_5<0)$, see Eq. (7.12) in \cite{y1}. When this criterion is applied to  \rn-like solutions, it reduces to $b_2>0$, which implies $k<0$. In the sector $k<0$ ($q<0$), which does not exist in the \rn\ case, we have generically just one horizon at $r=r_+$ ($r_-$ is negative), and the value of $p$ remains unrestricted. Thus, the standard \rn\ sector $q>0$, together with  possible naked singularities, is effectively eliminated from \rn-like spacetimes.

\subsection{Covariant momenta}

In our approach to black hole thermodynamics, one needs to have explicit expressions for the covariant momenta $H_i$ and $H_{ij}$. Starting from the Lagrangian \eq{3.12}, the covariant momenta for \rn-like black holes are found to be
\bea
&&H_i=2a_0\hd\Big(\ir{1}T_1-2\ir{2}T_i-\frac{1}{2}\ir{3}T_i\Big)\,,\nn\\
&&H_{ij}=-2a_0\hd(b_ib_j)+H'_{ij}\,,\quad
  H'_{ij}=2 b_2\hd\Big(\ir{2}R_{ij}-\ir{3}R_{ij}
                                   -\frac{1}{3}\ir{5}R_{ij} \Big)\,.\lab{3.13}
\eea
In more details,
\bsubeq
\bea
&&H_0=-H_1=\frac{2a_0p}{Nr}\vth^0\vth^1-2a_0\frac{N}{r}\vth^2\vth^3\,,\nn\\
&&H_2=-\frac{a_0}{Nr^3}(q-r^2)\vth^-\vth^3\, ,                       \nn\\
&&H_3=\frac{a_0}{Nr^3}(q-r^2)\vth^-\vth^2\, ,
\eea
\bea
&&H_{01}= -2a_0\vth^2\vth^3-\frac{2b_2p}{3r^2}\vth^0\vth^1\, ,\qquad
  H_{12}=-2a_0\vth^0\vth^3+\frac{b_2p}{3r^2}\vth^-\vth^2\,,          \nn\\
&&H_{02}=2a_0\vth^1\vth^3+\frac{b_2p}{3r^2}\vth^-\vth^2\,,\qquad
  H_{13}= 2a_0\vth^0\vth^2+\frac{b_2p}{3r^2}\vth^-\vth^3\,,          \nn\\
&&H_{03}=-2a_0\vth^1\vth^2+\frac{b_2p}{3r^2}\vth^-\vth^3\,,\qquad
  H_{23}=-2a_0\vth^0\vth^1-\frac{4b_2p}{3r^2}\vth^2\vth^3\, .      \lab{3.14b}
\eea
\esubeq

In addition to being used in thermodynamic analysis, the above results can also help us to understand the limit $p=0$ of \rn-like black holes. First, note that the piece $H'_{ij}$ of the covariant momentum \eq{3.13} is found to be linear in $p$, as shown in \eq{3.14b}. Hence, it vanishes in the limit $p=0$, and consequently, the quadratic curvature part of $L_G$, proportional to $R^{ij}H'_{ij}$, also vanishes. Then, using the well-known PG identity displayed in Eq. (5.9.18) of Ref. \cite{y2}, one can directly conclude that the remaining two parts of $L_G$ reduce, up to a divergence, to the GR Lagrangian. Thus:
\bitem
\item[--] In the limit $p=0$, the gravitational Lagrangian \eq{3.12}, associated to \rn-like black holes, reduces to the GR form; moreover, since $q\equiv kp^2=0$, a \rn-like spacetime reduces to the Schwarzschild spacetime of GR.
\eitem

\section{Black hole thermodynamics}\label{sec4}
\setcounter{equation}{0}

Basic elements of the general variational equation \eq{2.3} for the asymptotic charges and entropy are a Killing vector $\xi$ ($\pd_t$ in the present analysis), the dynamical variables $\vth^i$ and $\om^{ij}$, and the covariant momenta $H_i$ and $H_{ij}$.

\subsection{Energy}

The expression for energy $E_t$ is determined from the relation
$\d E_t:=\d\G_\infty$ for $\xi=\pd_t$, where only the parameters of the solution, $m$ and $q$ (or equivalently $p$), are varied. Using
\be
N\d N=-\frac{\d m}{r}+\frac{\d q}{2r^2}\,,   \nn
\ee
one finds the following nonvanishing contributions to $\d\G_\infty$ (integration over the boundary at infinity is implicit):
\bea
&&\vth^0{}_t\d H_0=-2a_0\frac{N\d N}{r}\vth^2\vth^3=8\pi a_0\d m\,,  \nn\\
&&\d\om^{12}H_{12t}=-2a_0\d\Big(\frac{N}{2r}\vth^2\Big)N\vth^3
                                          =4\pi a_0\d m\,,           \nn\\
&&\d\om^{13}H_{13t}=2a_0\d\Big(\frac{N}{2r}\vth^3\Big)N\vth^2
                                          =4\pi a_0\d m\,.           \nn
\eea
Then, summing up and using the normalization $16\pi a_0=1$, one obtains
\be
\d E_t\equiv\d\G_\infty=\d m\, .                                                 \lab{4.3}
\ee

\subsection{Entropy}

Relying on the relations valid on horizon
\be
N\pd_r N\big|_{r_+}=\k\,,\qquad N\d N\big|_{r_+}=0\,,                \nn
\ee
and for $\xi=\pd_t$, one finds the following nonvanishing contributions to $\d\G_H$
\bea
&&\d\vth^2H_{2t}=\d\vth^2a_0(2N')N\vth^3=8\pi a_0\k r_+\d r_+\,,     \nn\\
&&\d\vth^3H_{3t}=8\pi a_0\k r_+\d r_+\,,                             \nn\\
&&\om^{23}{}_t\d H_{23}
  =-\frac{16\pi b_2}{3}\frac p{r_+}\d p=16\pi a_0k\frac{p\d p}{r}\,, \nn
\eea
where the integration over the horizon is implicitly understood.  Thus,
\be
\d\G_H=\frac{1}{2}\k\d r_+^2+\frac{1}{2r_+}\d q\,.                   \lab{4.2}
\ee

The metric parameter $q=kp^2$ is induced by the torsion field which, together with curvature, represents an intrinsic feature of Riemann-Cartan geometry. However, the role of $q$ in the \rn-like metric is formally the same as the role of the electric charge in the standard \rn\ case. This similarity is reflected in the form of the second term in \eq{4.2}. Thus, the genuine gravitational parameter $q$ affects $\d\G_H$ as if it were the electric charge $Q$ of some external electromagnetic field. Similar thermodynamic roles of $q$ and $Q$ can be seen more explicitly by comparing our Eq. \eq{4.2} and Eq. (15) in Wald's work \cite{x3}$_1$.

Going now back to the first term of $\d\G_H$, one can conclude that it consistently defines entropy as a local function of thermodynamic variables \cite{x2},
\bea
T\d S=\frac{\k}{2}\d r_+^2\, ,\qquad S:=\pi r_+^2\,,
\eea
whereupon the final form of $\d\G_H$ reads
\be
\d\G_H=T\d S+\frac{1}{2r_+}\d q\,.                                   \lab{4.4}
\ee

\subsection{The first law}

Having found $\d \G_\infty$ and $\d\G_H$, we now wish to verify the validity of the first law,
\be
\d\G_\infty=\d\G_H\,.                                                \lab{4.5}
\ee
Using Eq. \eq{3.2}$_1$ for horizon and the expression \eq{3.3} for surface gravity, a simple algebraic manipulation yields a relation between the thermodynamic variables $(m,S,q)$, known as the Smarr formula \cite{x17},
\be
m=\k r_+^2+\frac{q}{r_+}\equiv 2TS+\frac{q}{r_+}\, .                 \lab{4.6}
\ee
The variation of this equation yields the identity \cite{x18}
\be
\d m=\frac{1}{2}\k\d r_+^2+\frac{1}{2r_+}\d q\,,                     \lab{4.7}
\ee
which proves the validity of the first law \eq{4.5}.

\section{Concluding remarks}\label{sec5}

In the present paper, we applied the Hamiltonian approach proposed in Ref. \cite{x5} to analyze energy, entropy and the first law of the \rn-like black hole with torsion, found in the dynamical framework of PG \cite{x1}, see also \cite{x12}.

The Riemann-Cartan geometry of \rn-like spacetimes is defined by the tetrad field (or metric) and the Lorentz connection. The \rn-like metric is obtained from the standard \rn\ metric by replacing the electric charge of the external electromagnetic source with a genuine gravitational parameter $q$. Technically, $q$ is produced by the torsion parameter $p$ through the relation $q=k\,p^2$. In contrast to the electric charge, $q$ is not a conserved quantity but just a \emph{hair parameter} of the solution. The condition $b_2>0$, necessary for the positivity of energy in the spin 1 sector of PG, implies that $q$ has to be negative.

Entropy of the black hole is determined by the variation of the boundary term at horizon, $\d\G_H$. The first part of $\d\G_H$ defines entropy as a local function of the thermodynamic variables, given by one-fourth of the horizon area,
$S=\pi r_+^2$, whereas the second part is expressed by a contribution stemming from the hair parameter $q$. Although the first law looks like the corresponding law in GR, its dynamical content is quite different.

Using the present approach, one can show that energy and entropy of a slightly extended \rn-like black hole found by the same authors \cite{x19}, coincide with the results obtained here.

\section*{Acknowledgments}
\vsm
We would like to thank Yuri Obukhov for useful comments.
This work was partially supported by the Ministry of Education, Science and Technological Development of the Republic of Serbia.

\end{document}